\documentclass[%
reprint,
 amsmath,amssymb,
 aps,
superscriptaddress,
]{revtex4-1}

\usepackage{graphicx}
\usepackage{dcolumn}
\usepackage{bm}
\usepackage{color,soul} 
\usepackage{upgreek}

\begin{document}

\preprint{APS/123-QED}

\title{Dislocation Networks and the Microstructural Origin of  Strain Hardening}

\author{Ryan B. Sills}
 \email{Corresponding author: rbsills@sandia.gov}
\affiliation{
 Sandia National Laboratories, Livermore, CA 94551
}%
\affiliation{
 Department of Mechanical Engineering, Stanford University,  Stanford, CA 94305
}%

\author{Nicolas Bertin}
\author{Amin Aghaei}
\author{Wei Cai}%
 \email{caiwei@stanford.edu}
\affiliation{
 Department of Mechanical Engineering, Stanford University,  Stanford, CA 94305
}%

\date{\today}


\begin{abstract} 
When metals are plastically deformed, the total density of dislocations
increases with strain as the microstructure is continuously refined, leading to
the strain hardening behavior.
Here we report the fundamental role played by the junction formation process in
the connection between dislocation microstructure evolution and the strain
hardening rate in face-centered cubic (FCC) Cu, as revealed by discrete
dislocation dynamics (DDD) simulations.
The dislocation network formed during $[0\,0\,1]$ loading consists of line
segments whose lengths closely follow an exponential distribution.
This exponential distribution is a consequence of junction formation by
dislocations on different slip planes, which can be modeled as a
one-dimensional Poisson process.
We show that, according to the exponential distribution, the dislocation
microstructure evolution is governed by two non-dimensional parameters, and the
hardening rate is controlled by the rate of stable junction formation.
%
%
%
By selectively disabling specific junction types in DDD simulations, we find 
that among the four types of junctions in FCC crystals, glissile junctions make 
the dominant contribution to the strain hardening rate.
\end{abstract}

\pacs{Valid PACS appear here}
\maketitle




The flow stress required to continuously deform a crystal generally increases
with the amount of plastic strain; this phenomenon is called strain hardening.
The strain hardening rate is one of the most prominent features of the
stress-strain curves of materials, and is critical for the stability of plastic
flow against local necking~\cite{Ghosh1977}. 
A quantitative understanding of the strain hardening rate in terms of
fundamental physical mechanisms has attracted significant interest not only as a
challenging problem in non-equilibrium statistical
mechanics~\cite{Friedel1955,Seeger1957,Nabarro1964,Mecking1981,Kocks2003,Bulatov2006,Devincre2008},
but also for its importance in advanced manufacturing
processes~\cite{Hosford2011B} (e.g., forming and cold working) and novel alloy
design~\cite{Tien1976B}.

At temperatures below about one-third of the melting point of a metal, 
movement of dislocations---line defects in the crystal lattice---is the
dominant mechanism for plastic deformation. 
It is widely believed that under such conditions, the strain hardening behavior
of pure, single crystalline metals is entirely governed by the dynamics of 
dislocations, which multiply and form intricate network structures during 
plastic deformation.
Through extensive theoretical and experimental research over the last five
decades~\cite{AHL2017},
%
%
a great deal is now known regarding the dislocation processes and strain
hardening behaviors of single crystals. The mobility of individual dislocations
and reactions between them are well understood through elasticity
theory~\cite{Shenoy2000,Madec2002,Madec2003,Kubin2003,Devincre2006,Bulatov2006}
and atomistic
simulations~\cite{Rodney1999,Bulatov2002,Olmsted2004b,Weinberger2011}.
Dislocation microstructures have been extensively characterized using
transmission electron microscopy \cite{Basinski1979,Hansen1995,Hong2013}. The
stress-strain curves for single crystals have also been measured under a wide
range of temperatures and loading directions \cite{Honeycombe1984B}.
However, a quantitative connection between the key microstructural features of
the dislocation network and the strain hardening rate of the metal is still
lacking.

In principle, the missing connection can be provided by large-scale discrete
dislocation dynamics (DDD) simulations, which follow the evolution of the
dislocation network and predict the stress-strain curve of the crystal.  
Using the {\tt ParaDiS} program~\cite{Arsenlis2007,ParaDiS} and a recently
developed time integrator~\cite{Sills2016a}, we can now predict the
stress-strain curves of single crystal Cu along the $[0\,0\,1]$ direction to a
sufficient amount of strain so that the strain hardening rate can be obtained
consistently.
Our DDD simulations reveal an important microstructural feature of the
dislocation network: the lengths of dislocation line segments (between junction
nodes) closely obey an exponential distribution.
This exponential distribution is parameterized by the dislocation density
$\rho$, and a dimensionless parameter $\phi = N^2/ \rho^3$, where 
$N$ is total number of the dislocation line segments per unit volume.
We show that the exponential length distribution is the result of a
one-dimensional Poisson process, where a dislocation line segment is randomly
subdivided into two shorter segments when a stable junction is formed.
Furthermore, junction formation is found to be essential for dislocation multiplication and
strain hardening, because the strain hardening rate vanishes when
junction formation is disabled in DDD simulations.
Based on the exponential distribution of line lengths, a quantitative connection
is established between the junction formation rate and the strain hardening
rate.
And, by selectively disabling different junction types in DDD simulations, we find
that glissile junctions make the dominant contribution to the strain hardening
rate.
%
%


Our DDD simulation cell is a cube with a $15\,\upmu{\rm m}$ length subjected to
periodic boundary conditions in all three directions. 
The initial condition consists of straight dislocation lines on $\{1\,1\,1\}$
planes with $\frac{1}{2}\langle 0\,1\,1\rangle$ Burgers vectors. These
dislocations are confined to glide on $\{1\,1\,1\}$ planes. 
%
A linear mobility is applied to all mobile dislocations with a drag coefficient
of $B = 15.6\,\upmu$Pa$\cdot$s. 
The elastic interactions between dislocations are parameterized by the
shear modulus $\mu = 54.6$~GPa and Poisson's ratio $\nu = 0.324$ of Cu. 
The dislocation structure is first relaxed to an equilibrium state, giving the
initial density $\rho_0$, and is then subjected to a constant strain rate
$\dot{\varepsilon}$ along $[0\,0\,1]$. 
%
%
Under this high-symmetry loading condition, eight of the twelve slip systems 
have the same resolved shear stress with Schmid factor $S\approx0.41$, and 
the remaining four have a Schmid factor of zero.
%

\begin{figure}
\center
\includegraphics[height=2.5in]{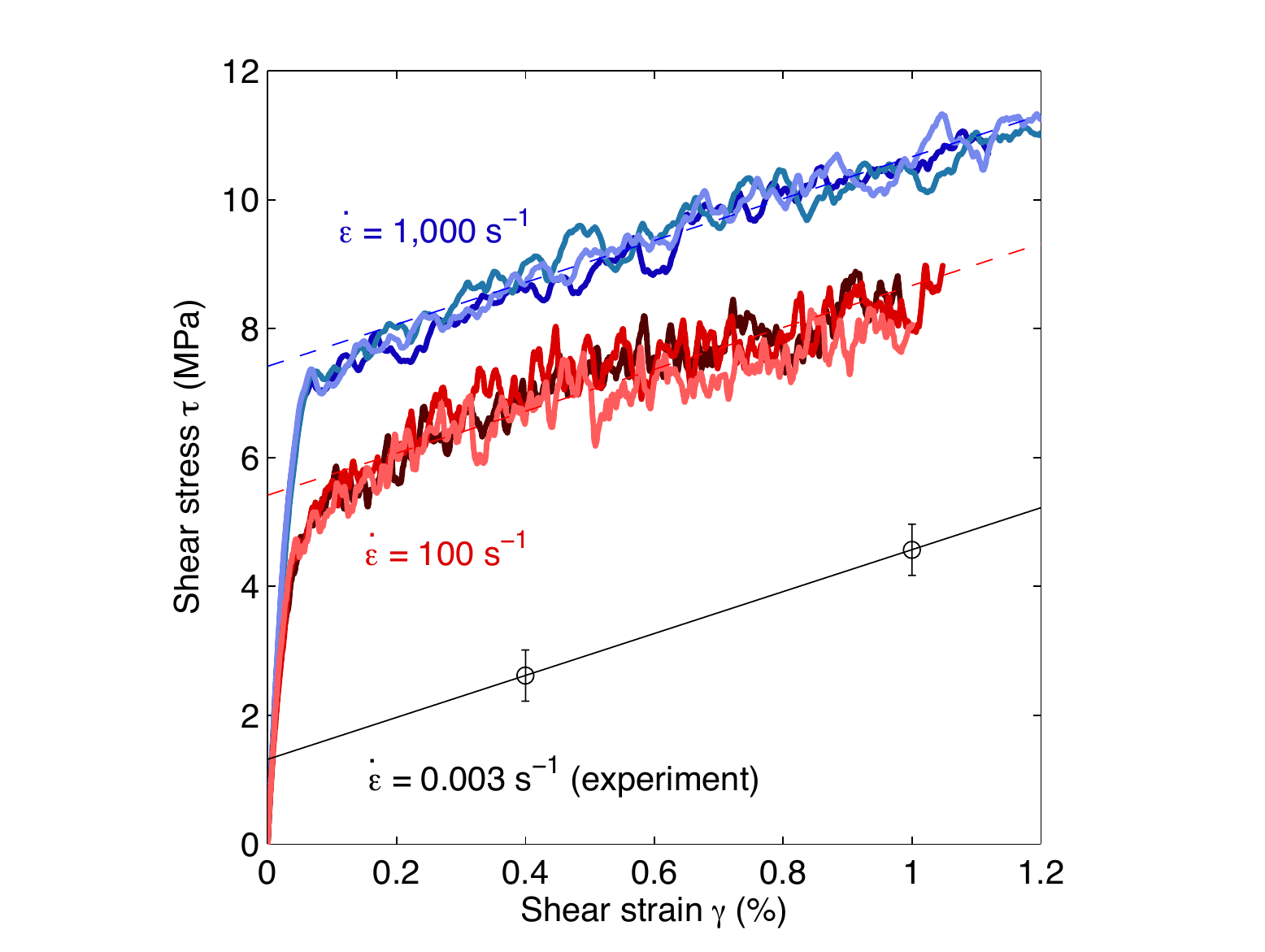} 
%
\caption{\label{fig:stressstrain} Shear stress-strain curves of single
crystalline Cu deformed along $[0\,0\,1]$ axis.
Thick curves are predictions by DDD simulations for an initial dislocation
density $\rho_0 = 0.7 \times 10^{12}$ m$^{-2}$ at two different strain rates
$\dot{\varepsilon}$.
%
%
%
The thin solid line is extracted from experimental data~\cite{Takeuchi1975} at
strain rate of $\dot{\varepsilon} = 3\times 10^{-3}$ s$^{-1}$.
The dashed lines are translated from the thin solid line to show the consistency
of the strain hardening rate between simulations and experiments.
%
%
%
%
%
}
\end{figure}

Fig.~\ref{fig:stressstrain} shows the shear stress-strain curves predicted by
DDD simulations under two strain rates, where the shear stress $\tau$ and strain
$\gamma$ are related to normal stress $\sigma$ and strain $\varepsilon$ by 
$\tau = S\sigma$ and $\gamma = \varepsilon/S$. 
Three independent simulations with randomly generated initial
conditions are performed for each strain rate and initial density to insure
robustness of the results ({additional results with different initial densities} are in
the Supplementary Information).
%
%
The strain hardening rate is largely consistent with the Stage II 
hardening rate of $\Theta \equiv d\tau/d\gamma \sim$ 320$\pm 50$~MPa observed by
Takeuchi~\cite{Takeuchi1975} in single crystalline copper, and the commonly
cited rules-of-thumb of $\mu/200$ to $\mu/300$ (180-270~MPa) for FCC metals in
Stage II \cite{Basinski1979,Kocks2003}.
The consistency in the slope of the stress-strain curves in
Fig.~\ref{fig:stressstrain} shows that the strain hardening rate is less
sensitive to the applied strain rate than the yield stress. 
%

%
%

The success of DDD simulations in capturing the strain hardening rate of single
crystals enables us to answer the fundamental question: which microstructural
features in the dislocation network are responsible for the strain hardening
behavior?
An obvious candidate is the total dislocation density $\rho$.  In fact, the 
well-known Taylor relation states that the flow stress $\tau$ satisfies the 
following relation,
\begin{equation} 
  \tau = \alpha\, \mu \, b \, \sqrt{\rho} \label{eq:taylor}
\end{equation} 
where $b$ is the magnitude of the Burgers vector, and $\alpha$ is a constant
experimentally determined to be between $0.5$ and $1$ \cite{Mecking1981}. 
We find that the Taylor relation is obeyed in our DDD simulations during hardening
with $\alpha \approx 0.5$ (see Supplementary Information). 

%

\begin{figure}
\center
 \includegraphics[height=2.1in]{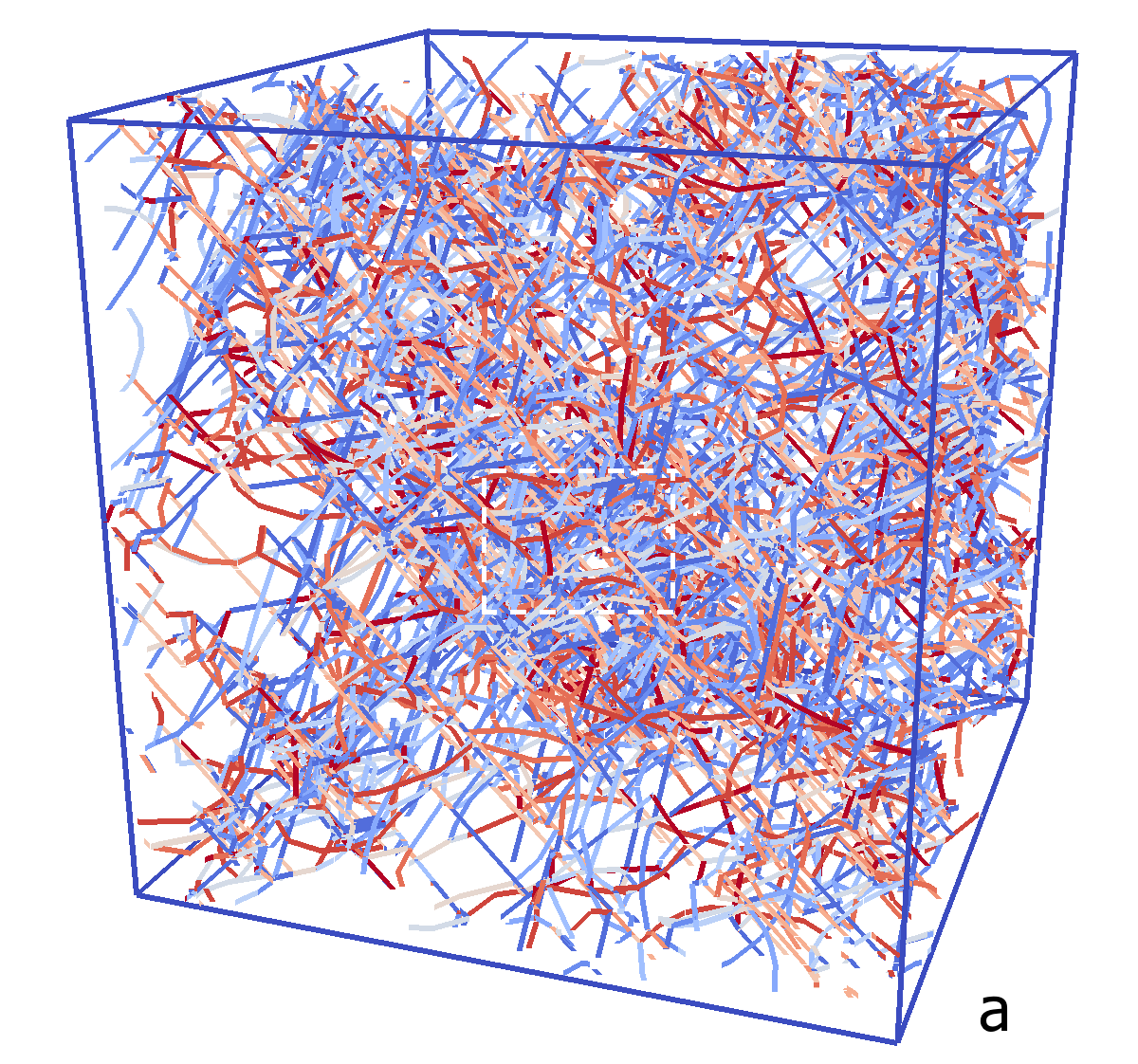} \\ 
 \includegraphics[height=2.3in]{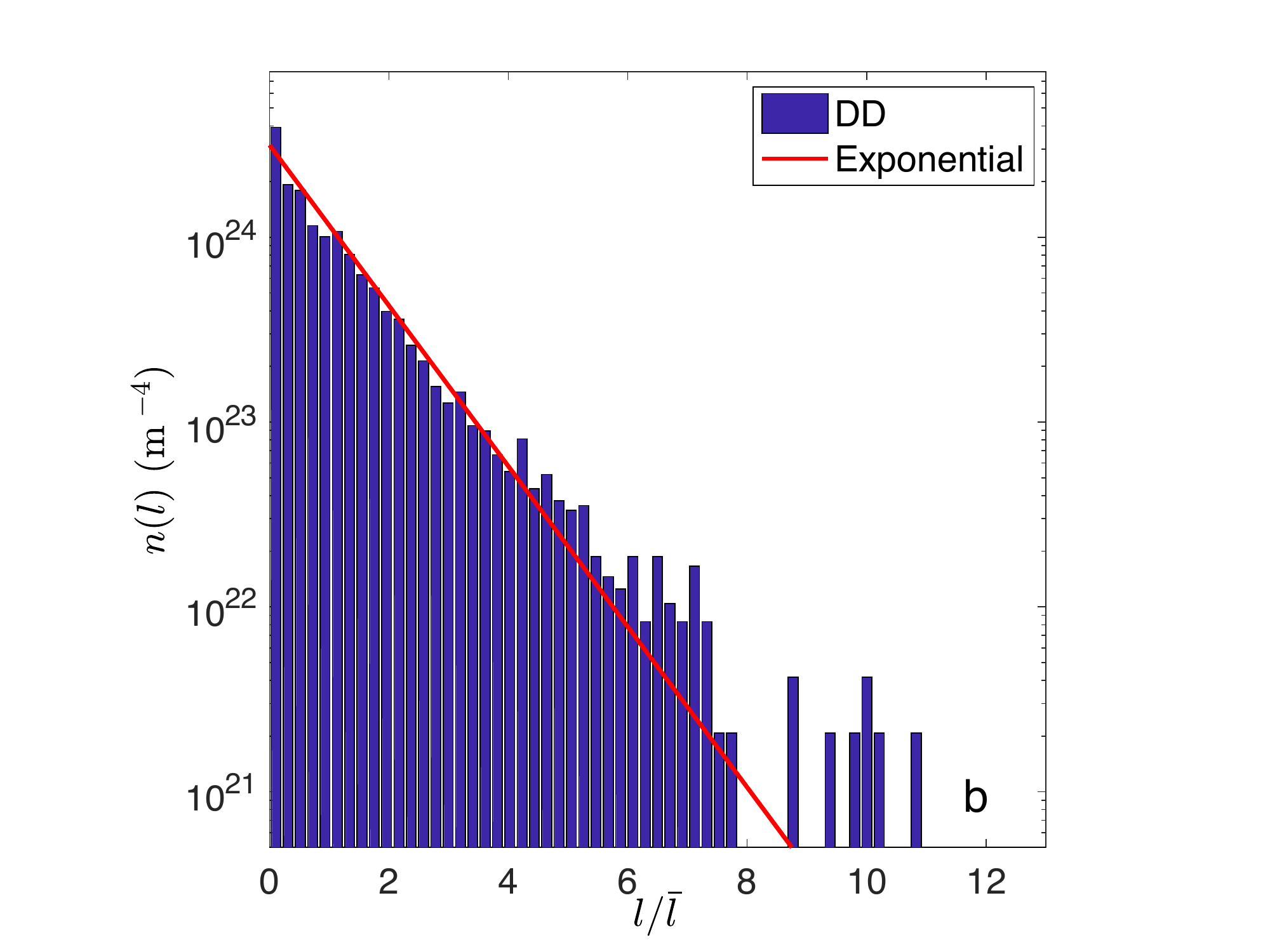} 
\caption{\label{fig:snapshot} (a) Snapshot from a DDD simulation at $\gamma =
0.87\%$ shear strain, with $\dot{\varepsilon}=10^3$ $\rm s^{-1}$ and $\rho_0 =0.7
\times 10^{12}$ m$^{-2}$. (b) Dislocation link length distribution $n(L)$ for
the structure shown in (a) (after being relaxed to zero stress) compared to
Eq.~(\ref{eq:nLexponential}).}
\end{figure}

%
The Taylor relation has often been interpreted by considering the critical
stress needed to free (or activate) dislocations whose end points are pinned
(e.g., by junctions)~\cite{Saada1960}, by assuming that the average segment length 
scales with $\rho^{-1/2}$.
Fig.~\ref{fig:snapshot}(a) shows a typical dislocation microstructure from our
DDD simulations. The dislocation line network is quite complex and the 
segment lengths are clearly not all the same as is often assumed in simple
models. 
Given the fundamental connection between dislocation line length and flow
stress, the distribution of line lengths should be of primary importance when
characterizing the statistical properties of dislocation networks, although in
the past it has not received much attention.
%
%
%
In this work, we have analyzed the length of all dislocation lines connecting 
one pinning point (e.g., nodes terminating at junctions) to another, which has 
been referred to as the \emph{link length}~\cite{Bilde-Sorensen1973}.
We have performed such an analysis both on instantaneous snapshots from DDD
simulations, as well as on configurations after the snapshots are relaxed under
zero stress (similar to postmortem analysis in experiments).

Fig.~\ref{fig:snapshot}(b) shows the distribution of link lengths $l$
normalized by the average length $\overline{l}$ from a typical relaxed 
dislocation structure at $\gamma = 0.87$\% shear strain.
The histogram shows that the probability distribution of link lengths can be 
well described by an exponential distribution,
\begin{equation}
  p(l) = (1 /\overline{l} ) \, {\rm e}^{-l / \overline{l} }.  \label{eq:plexp}
\end{equation}
Other researchers have found that dislocation networks observed in
molecular dynamics simulations of strained nanopillars
exhibit a similar exponential distribution~\cite{Voyiadjis2017}.
For the distribution shown in Fig.~\ref{fig:snapshot}(b), we find that for link
lengths less than $2.5\,\overline{l}$ (which comprises 91\% of the links), the
deviation from the exponential distribution in a quantile-quantile (Q-Q) plot is
less than 5\% (see Supplementary Information).
%

\begin{figure}
\center
 \includegraphics[height=2.3in]{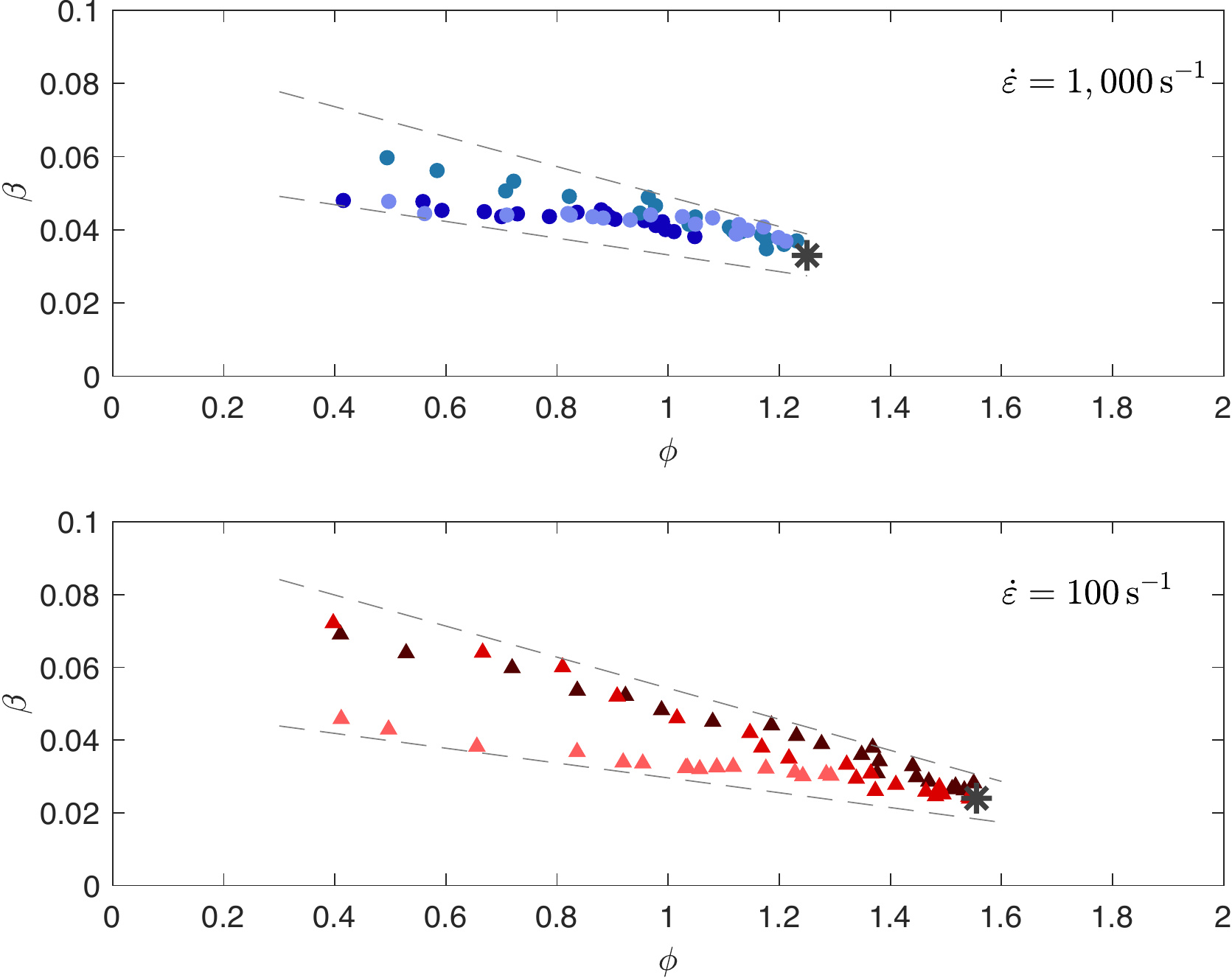}
%
\caption{\label{fig:beta_phi} $\phi$-$\beta$ values during DDD simulations
at strain rates $\dot{\varepsilon} = 10^3\, {\rm s}^{-1}$ and $10^{2}\,{\rm s}^{-1}$.
%
%
Dashed lines are guide to the eye for the range of $\phi$-$\beta$ values.
The $*$ sign indicates the critical value $\phi_{\rm c}$, beyond which
Eq.~(\ref{eq:Theta_beta_simp}) becomes valid.
}
\end{figure}

To discuss the nature of the exponential length distribution and its consequence
on strain hardening, it is useful to introduce a length density function, $n(l)
\equiv N\,p(l)$, where $n(l)\,dl$ is the number of links per unit
volume whose length is between $l$ and $l+dl$.
$N$ is the total number of links per unit volume.
$N$ and $\rho$ are the zeroth and first moments of the density function $n(l)$,
respectively.  
If we define a non-dimensional parameter $\phi \equiv N^2 / \rho^3$, then the
average link length is $\overline{l} = \rho / N = 1 / \sqrt{\phi\,\rho}$, and 
Eq.~(\ref{eq:plexp}) is equivalent to
\begin{equation}
   n(l) = \phi\, \rho^2 \, {\rm e}^{ -\sqrt{\phi\,\rho}\, l}.  \label{eq:nLexponential}
\end{equation}
Eq.~(\ref{eq:nLexponential}) means that the dislocation length distribution is
completely determined by the total density $\rho$ and the non-dimensional
parameter $\phi$. 
We find that $\phi$ gradually increases during the course of the simulations (see 
Fig.~\ref{fig:beta_phi}).
%


The exponential form of $n(l)$ can be explained by considering a simple model 
for the junction formation process.
The degrees of freedom of the model are the length $l_i$ of $N$ dislocation links in a unit 
volume.
When a link of length $l$ participates in a junction formation
process, we assume that this link splits into two links of lengths $l_1$ and
$l_2 = l - l_1$, where $l_1$ is uniformly distributed in $(0,l)$.
If we assume that the probability rate for splitting a link is proportional to
its length $l$, then numerical simulations show that the distribution of
link lengths quickly goes to an exponential distribution (see Supplementary
Information).
Therefore, we attribute the origin of the link length distribution to the
observation that the probability of a link of length $l$ experiencing a
collision is proportional to $l$, rendering junction formation a one-dimensional
Poisson point process from which an exponental distribution
results~\cite{Lanchier2017B}.
In other words, the probability of finding a link of length $l$ having not yet 
suffered a collision is exponentially small ($\sim {\rm
e}^{-l/\overline{l}}$).

In the following, we show that focusing on the dislocation link length distribution
function, $n(l)$, allows us to reveal deeper connections between the dislocation
microstructure and the strain hardening rate that have not been appreciated before.
We shall assume that $n(l)$ (in single crystal Cu under $[0\,0\,1]$ loading)
always follows the exponential distribution of Eq.~(\ref{eq:nLexponential}),
which is parameterized by the dislocation density $\rho$ and a non-dimensional
parameter $\phi$.
%
Furthermore, we shall assume that the dislocation microstructure (under
$[0\,0\,1]$ loading) is uniquely determined by the two parameters $\rho$ and
$\phi$.
Given these assumptions, we will be able to establish a quantitative link 
between the junction formation rate and the strain hardening rate, as shown 
below.

Consistent with the assumption that the probability rate of a line splitting is
proportional to its length $l$, we assume that the overall collision rate
between dislocations on different slip systems is $R_{\rm int} =
f\,\rho^2\,\overline{v}$, where 
$f$ is the ratio of the forest density $\rho_{\rm f}$ over the total density
$\rho$, and the mean spacing between pinning points on dislocation slip planes
is $\lambda \equiv 1/\sqrt{f\,\rho}$.
Our DDD simulations show that $f \approx 0.55$ remains a constant with
strain (see Supplementary Information).
The difference between $\phi$ (increasing with strain) and $f$ (constant)
means that the average link length $\overline{l}$ is
different from $\lambda$.
$\overline{v}$ is the average dislocation velocity related to the strain rate
$\dot{\gamma}$ through the Orowan equation, $\dot{\gamma} =
\rho\,b\,\overline{v}$.
%
%
%
Since not all collisions result in the formation of a stable junction,
we introduce a non-dimensional parameter $\beta$, which is the fraction
of collisions that leads to stable junction formation, so that the junction
formation rate is $\beta R_{\rm int}$.
We assume that each time a stable junction forms, two dislocation links become 4 
links, so that the rate at which the number of links increases is
\begin{equation}
 \dot{N} = 2 \, \beta\,f\,\rho^2\,\overline{v} 
             = \frac{2\,\beta\,f\,\rho}{b}  \, \dot{\gamma}.
 \label{eq:Ndot}
\end{equation}
Assuming that the exponential distribution is always maintained, it follows that 
$\dot{N} = \frac{3}{2}\phi^{1/2}\rho^{1/2}\dot{\rho}+
\frac{1}{2}\phi^{-1/2}\rho^{3/2}\dot{\phi}$, leading to the dislocation 
multiplication rate
\begin{equation}
\dot{\rho} = \frac{4\,\beta\,f\,\dot{\gamma}}{3\sqrt{\phi}\,b}\, \sqrt{\rho} 
   - \frac{\dot{\phi}}{3\,\phi}\rho.
  \label{eq:rho_dot}
\end{equation}
Interestingly, the dislocation multiplication rate expression in
Eq.~(\ref{eq:rho_dot}) is consistent with the Kocks-Mecking model~\cite{Kocks2003},
$\dot{\rho} = K_1 \sqrt{\rho} - K_2 \rho$, with 
$K_1 =  \frac{4\,\beta\,f\,\dot{\gamma}}{3\,\sqrt{\phi}\,b}$
and $K_2 = \frac{\dot{\phi}}{3\,\phi}$.
Here we see that the Kocks-Mecking form appears as a natural consequence of the 
exponential distribution of link lengths.
Combining Eq. ~(\ref{eq:rho_dot}) with the Taylor relation,
Eq.~(\ref{eq:taylor}), the strain hardening rate is,
\begin{equation}
  \Theta = \frac{\dot{\tau}}{\dot{\gamma}} = \frac{1}{3}\alpha\mu \, \left(
      \frac{2\,\beta\,f}{\sqrt{\phi}}
      - \frac{b\sqrt{\rho}}{2\,\phi} \frac{\dot{\phi}}{\dot{\gamma}} \right).
      \label{eq:Theta_beta}
\end{equation}
Eq.~(\ref{eq:Theta_beta}) shows that the strain hardening rate $\Theta$ is
determined by the junction formation rate $\beta$, as well as
$\dot{\phi}/\dot{\gamma}$, the rate at which parameter $\phi$ changes with
strain $\gamma$.

We can analyze the different contributions to the strain hardening rate $\Theta$ 
in our DDD simulations in light of Eq.~(\ref{eq:Theta_beta}).
Specifically, we compute the dislocation density $\rho$ and link number density $N$ 
from a series of DDD simulation snapshots, and extract the non-dimensional 
parameters $\phi$ and $\beta$ using
$\phi = N^2 / \rho^3$
and $\beta = {b\,\dot{N}}/({2\,f\,\rho\,\dot{\gamma}})$.
Given the assumption that the dislocation microstructure (for Cu in $[0\,0\,1]$
loading) is completely determined by $\rho$ and $\phi$, we expect the 
junction formation rate $\beta$, being a property of the dislocation network, to 
be a function of $\rho$ and $\phi$.  
Hence, for simulations starting from the same initial dislocation density
$\rho_0$ and at the same strain rate $\dot{\gamma}$, we expect $\beta$
(on average) to be a function of $\phi$.

Fig.~\ref{fig:beta_phi} shows the evolution of dislocation microstructure in the
non-dimensional space of $\beta$-$\phi$.
It is observed that during the course of the simulation, $\phi$ gradually 
increases from about 0.4 to as high as about 1.6, accompanied by a decrease in $\beta$.
This indicates that during this early period of strain hardening, the 
dislocation microstructure does not evolve in a self-similar manner---not only does
the density $\rho$ increase, but the non-dimensional parameter $\phi$ also 
increases.
However, the rate of increase of $\phi$ slows down with increasing strain, so
that if $\phi$ exceeds a critical value $\phi_{\rm c}$,
$\dot{\phi}/\dot{\gamma}$ becomes so small that the second term in
Eq.~(\ref{eq:Theta_beta}) becomes negligible compared to the first term (see
Supplementary Information).
%
In this case, the strain hardening rate can be approximated as
\begin{equation}
  \Theta \approx \frac{2\, \alpha\, \beta\,f}{3\,\sqrt{\phi}} \,  \mu .  \quad 
  \quad (\phi \ge \phi_{\rm c})
      \label{eq:Theta_beta_simp}
\end{equation}
%

Eqs.~(\ref{eq:Theta_beta}-\ref{eq:Theta_beta_simp}) motivate the hypothesis that
the net junction formation fraction $\beta$ is the controlling factor of the
dislocation multiplication rate and strain hardening rate.
In other words, if the dislocations were not allowed to form more junctions, the 
network would not have the capacity to store more dislocations and there would 
not be any strain hardening.
In order to test this hypothesis, we employed specialized DDD simulations in
which junction formation events are suppressed.
With our approach, we are able to suppress selected types of junctions,
enabling us to assess the role of each junction type (collinear, glissile,
Hirth, and Lomer \cite{Hirth1961,Kubin2013B}).
%
This is accomplished by preventing multi-node splits that result in junctions
\cite{Bulatov2006B,Arsenlis2007} 
and increasing the core radius to weaken short range elastic interactions
(see Supplementary Information).
%
%
This approach illustrates the power of computational tools such as DDD in
answering fundamental questions by examining scenarios not accessible by
experiments.

\begin{figure}
\center
\includegraphics[height=2.5in]{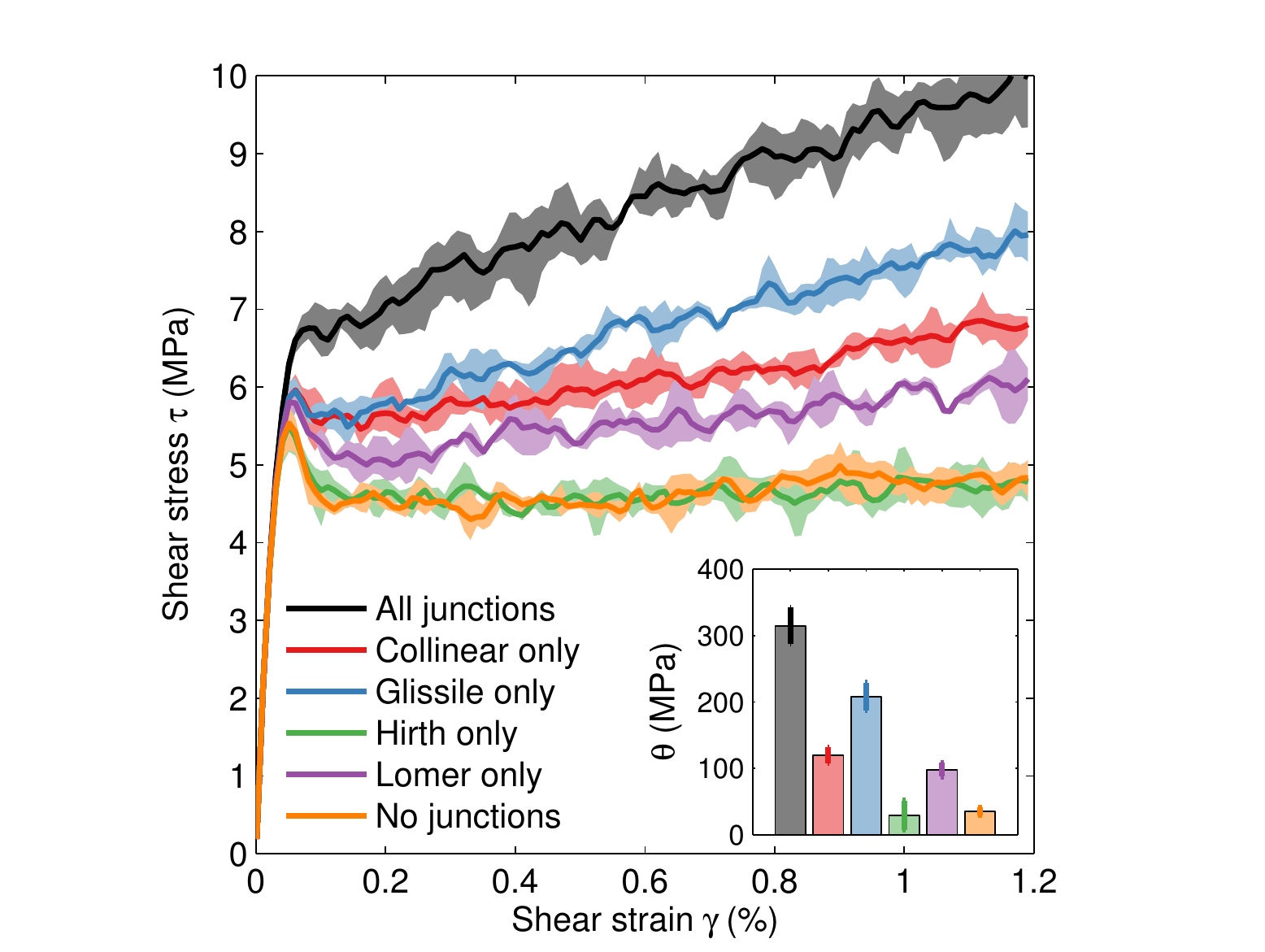}
%
\caption{\label{fig:junctionhardening} Shear stress-strain curves and (inset)
strain hardening rates for specialized DDD simulations with selected types of
junctions suppressed; $\rho_0 = 0.7 \times 10^{12}$ $\rm m^{-2}$ and
$\dot{\varepsilon} = 10^3$ $\rm s^{-1}$. 
The legend shows the type of junctions that are allowed to form during the
simulation. }
\end{figure}

Fig.~\ref{fig:junctionhardening} shows the stress-strain curves obtained with
these specialized simulations. When all junctions are suppressed, the hardening
rate (at $\dot{\varepsilon} = 10^{3}\,{\rm s}^{-1}$) is greatly reduced from 329
to 47~MPa. 
We believe the strain hardening rate does not completely vanish in this case 
because our modifications do not completely eliminate junction formation 
events.
When only one type of junction is allowed to form, the hardening rate
is approximately {225, 117, 103, and 46~MPa} for
glissile, collinear, Lomer, and Hirth junctions, respectively.
This result indicates that glissile junctions make the dominant contribution to 
strain hardening.  This finding is interesting because the collinear junction is 
the strongest among the four junctions~\cite{Madec2003}.
%
%
We believe that glissile junctions make the largest contribution because they
are relatively stable~\cite{Kubin2003} and are the most likely to form, with four forest
interactions per slip system (compared to two, two, and one for Lomer, Hirth,
and collinear, respectively).

%
Our DDD simulations have shown that the dislocation link lengths satisfies an
exponential distribution, that junction formation is a necessity for strain
hardening, and that glissile junctions make the dominant contribution to the
strain hardening rate in $[0\,0\,1]$ loading of Cu.
The exponential distribution is explained by a simple model for the effect
of  junction formation on the population of dislocation lines.
Analysis using the exponential length distribution reveals a
fundamental connection between the junction formation rate ($\beta$) and the
strain hardening rate ($\Theta$).
We hope this work brings attention to the dislocation link length distribution
as an important property of the dislocation microstructure, so that it may be
incorporated in future coarse-grained field theories of dislocation dynamics and
dislocation-based crystal plasticity models.

\acknowledgements 

This work was supported by Sandia National Laboratories (R.B.S.) and by the U.S. Department
of Energy, Office of Basic Energy Sciences, Division of Materials Sciences and Engineering under
Award No. {DE-SC0010412} (N.B., A.A. and W.C.).
Sandia National Laboratories is a multimission laboratory managed and operated
by National Technology and Engineering Solutions of Sandia, LLC, a wholly owned
subsidiary of Honeywell International, Inc., for the U.S. Department of Energy’s
National Nuclear Security Administration under contract {DE-NA0003525}.

\bibliography{wh_expdist_junc}


\end{document}